\documentclass[%
  reprint,
  amssymb,
  aps,
  prl,
  floatfix,
  superscriptaddress
]{revtex4-2}

\usepackage{silence}  
\usepackage{amsmath}
\usepackage{booktabs}
\usepackage{comment}
\usepackage{glossaries}
\usepackage{graphicx}
\usepackage[version=4]{mhchem}
\usepackage{siunitx}
\usepackage[dvipsnames]{xcolor}
\usepackage{hyperref}
\usepackage{tabularx}
\DeclareSIUnit\angstrom{\text {Å}}

\graphicspath{{./figures/}}
\setacronymstyle{long-short}
\newacronym[longplural={densities of state}]{dos}{DOS}{density of states}
\newacronym{gllbsc}{GLLB-sc}{Gritsenko-van Leeuwen-van Lenthe-Baerends-solid-correlation}
\newacronym{homo}{HOMO}{highest occupied molecular orbital}
\newacronym{hc}{HC}{hot carrier}
\newacronym{he}{HE}{hot electron}
\newacronym{hh}{HH}{hot hole}
\newacronym{ks}{KS}{Kohn-Sham}
\newacronym{lcao}{LCAO}{linear combination of atomic orbitals}
\newacronym{ld}{LD}{Landau damping}
\newacronym{lsp}{LSP}{localized surface plasmon}
\newacronym{lp}{LP}{lower polariton}
\newacronym{lumo}{LUMO}{lowest unoccupied molecular orbital}
\newacronym{np}{NP}{nanoparticle}
\newacronym{rttddft}{RT-TDDFT}{real-time time-dependent density functional theory}
\newacronym[longplural={projected densities of state}]{pdos}{PDOS}{projected density of states}
\newacronym{up}{UP}{upper polariton}
\newacronym{xc}{XC}{exchange correlation}

\hypersetup{colorlinks, 
	linkcolor={blue!75!black!80!yellow},
	citecolor={blue!75!black!80!yellow}, 
	urlcolor={blue!75!black!80!yellow}
}

\makeatletter \renewcommand\@make@capt@title[2]{%
\@ifx@empty\float@link{\@firstofone}{\expandafter\href\expandafter{\float@link}}%
\sffamily{\textbf{#1}}\@caption@fignum@sep#2 }

\usepackage[normalem]{ulem}

\newacronym{dft}{DFT}{density functional theory}
\newacronym{md}{MD}{molecular dynamics}
\newacronym{ml}{ML}{machine learning}
\newacronym{nep}{NEP}{neuroevolution potential}
\newacronym{pes}{PES}{potential energy surface}
\newacronym{qedft}{QEDFT}{quantum electrodynamical density-functional theory}

\newcommand{\gpaw}{GPAW}

\newcommand{\addphysics}{Department of Physics, Chalmers University of Technology, 412 96 G\"oteborg, Sweden}

\begin{document}

\newcommand{\papertitle}{Controlling Plasmonic Catalysis via Strong Coupling with Electromagnetic Resonators}

\hypersetup{pdfauthor={Jakub Fojt, Paul Erhart, and Christian Schäfer}}
\hypersetup{pdftitle={\papertitle}}

\title{\papertitle{}}
\author{Jakub Fojt}
\author{Paul Erhart}
\author{Christian Sch\"afer}
\email[Electronic address:\;]{christian.schaefer.physics@gmail.com}
\affiliation{\addphysics}

\date{\today}

\begin{abstract}
Plasmonic excitations decay within femtoseconds, leaving non-thermal (often referred to as ``hot'') charge carriers behind that can be injected into molecular structures to trigger chemical reactions that are otherwise out of reach -- a process known as plasmonic catalysis. 
In this Letter, we demonstrate that strong coupling between resonator structures and plasmonic nanoparticles can be used to control the spectral overlap between the plasmonic excitation energy and the charge injection energy into nearby molecules.
Our atomistic description couples real-time density-functional theory self-consistently to Maxwell's equations via the radiation-reaction potential.
Control over the resonator provides then an additional knob for non-intrusively enhancing plasmonic catalysis and dynamically reacting to deterioration of the catalyst -- a new facet of modern catalysis.
\end{abstract}


\maketitle

\Gls{hc} technology, i.e., injecting \glspl{hc} (non-thermal carriers) into a molecule~\cite{ZhoSweZha18} or semiconductor~\cite{GenAbdBer21}, promises considerable improvements in light-harvesting \cite{GenAbdBer21}, solar-to-chemical energy conversion \cite{AslRaoCha18, LiCheRic21, DuCTagWel18}, and catalysis \cite{ZhoLouBao21, DuCTagWel20, HouCheXin20, gupta2021nanoimprint, herran2023plasmonic, Yuan2024}.
Commonly, \glspl{hc} are generated in plasmonic \glspl{np} through the non-radiative decay of the \gls{lsp}, a mode of collective electronic motion that is excited by light.
This process is highly efficient due to the large absorption cross section of plasmonic \glspl{np} at visible-near UV frequencies \cite{Boh83, LanKasZor07}.
One possible process of injecting those generated \glspl{hc} into a molecule follows a \textit{direct} \gls{hc} transfer \cite{KhuPetEic21} where charge-transfer excitations form with one carrier in the \gls{np} and the other in the orbital of a molecule.
Such a direct process is more useful in terms of selectivity \cite{LinAslBoe15} and is at least as likely \cite{KhuPetEic21} as the process of transferring \glspl{hc} that are formed in the \gls{np} across the interface \cite{ZhoLouBao21}.
The direct \gls{hc} transfer process sensitively depends on the alignment of energetic levels comprising charge transfer and \gls{lsp} excitations \cite{FojRosKui22}.
Improving \gls{hc} generation and injection are therefore critical to explore the full potential of plasmonic catalysis.

One possible angle to improve plasmonic catalysis is to control the interplay between plasmonic particles and an optical field. 
Confining optical modes, may it be via structured meta-surfaces or Fabry-P\'erot cavities, results in an increase in interaction to a material with spectral overlap.
From an increase in mode density follows, according to Fermi's golden rule, an increase in photoabsorption cross section, which has been successfully employed to deposit more energy in plasmonic \glspl{np}, create more \glspl{hc}, and thus further increase catalytic efficiency \cite{gupta2021nanoimprint, Ren21,Jin23, HerJueKes23, NanGirSte23, Yuan2024}.
At sufficiently strong interaction, light and matter hybridize into polaritonic quasi-particles that can enhance exciton~\cite{feist2015, xiang2017twodim, du2018theory, groenhof2019tracking, deJong24} or charge~\cite{orgiu2015, herrera2016, schafer2019modification, Kumar24} conductance and even control chemical reactivity~\cite{hutchison2012, Munkhbat2018, ahn_herrera_simpkins_2022, chen2022cavity, galego2016, fregoni2020strong, schafer2021shining, Schaefer_2024}.
Plasmonic nanoparticle crystals \cite{mueller2020deep, Hertzog21} can reach extreme light-matter coupling strengths, entering the deep strong coupling domain, that even exceeds the excitation energy of the \gls{lsp} of the \glspl{np} comprising the crystal.

In this Letter, we explore to which extent strong coupling between an optical resonator and a plasmonic Ag \gls{np} can be leveraged to control the catalytic effect on a nearby CO molecule.
In contrast to previous works, our approach is based on energetic restructuring, i.e., each step in the catalytic process is associated with a set of energies and the efficiency of the full process is optimized by aligning those energies.
Using an atomistic description of the \gls{np}-molecule system, we show that the microscopic mechanism of charge injection to the molecules is a dephasing of the \gls{lsp} to charge-transfer excitations. \cite{RosKuiPus17, RosErhKui20, FojRosKui22}
We then couple the system to an optical cavity \cite{schafer2021shortcut}, giving rise to polaritonic states that emerge from hybridization of light and matter.
The polaritonic states allow tuning of previously mismatched energies, providing non-intrusive control that increases the efficacy of \gls{hc} injection into the molecule.
We conclude with a comprehensive discussion of potential applicability and limitations, including a comparative study for engineering the shape of the \gls{np}.

\begin{figure*}[ht]
    \includegraphics{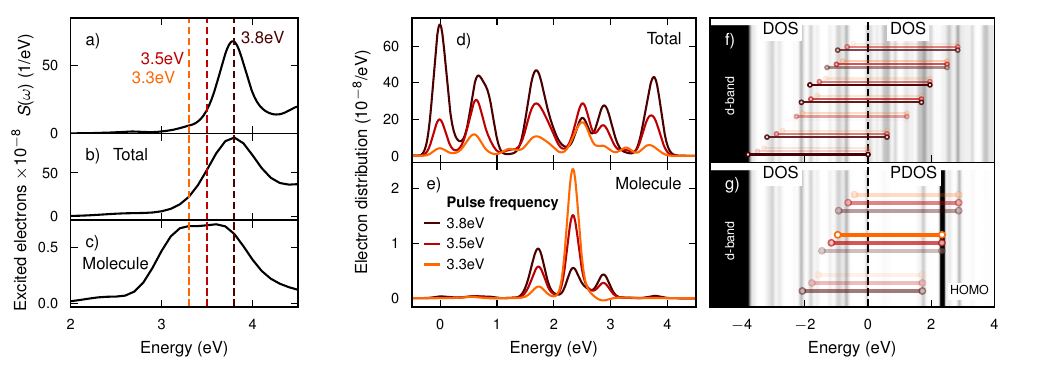}
    \caption{
    (a) Absorption spectrum of the \gls{np}-CO coupled system. The colours of the vertical dashed lines correspond to 3 exemplary excitation energies.
    (b) Number of excited electrons in the total system, and (c) in the molecule.
    (d) Electron distribution in the total system, and (e) projected on the molecule.
    (f) Schematic illustration of the energy-conserving excitations that contribute to excited electrons in total, overlaid on the total \gls{dos}.
    As there are many available states, there are many possibilities to satisfy the energy conservation condition (see text). 
    (g) Schematic illustration of the energy-conserving excitations that contribute to excited electrons in the molecule, overlaid on the total \gls{dos} of unoccupied states and \gls{pdos} of the molecule of unoccupied states.
    The sparse \gls{pdos} defines much stricter resonance conditions such that energetic matching becomes an essential component of catalytic efficiency.
    The opacity of the transitions in (f-g) is proportional to the relative probability of each particular excitation.
    }
    \label{fig:transfer-no-cavity}
\end{figure*}

\textit{System --}
Our exemplary model system comprises a CO molecule near a 201-atom Ag \gls{np} (effectively \qty{1.5}{\nano\meter} in diameter).
The \gls{lsp} of the \gls{np} (resonance \qty{3.8}{\electronvolt}; see \autoref{fig:transfer-no-cavity}a) and molecular excitations have no spectral overlap.
Carriers that form on the molecule are then only due to a photocatalytic effect of the excited \gls{lsp} dephasing into charge-transfer excitations -- providing unambiguous insight into the \gls{hc} injection.
We place the molecule \qty{3}{\angstrom} from the (111) face of the \gls{np}, as the alignment is most clearly illustrated here, but discuss other distances in the SI.
A detailed description of methodology and dephasing process can be found in the SI and Refs.~\citenum{RosErhKui20, FojRosKui22}.

\textit{Hot-carrier transfer --}
A prerequisite for the generation of \glspl{hc} is the absorption of energy.
It is therefore not surprising that when varying the frequency of the driving pulse, the number of excited electrons in the system (i.e., \gls{np} and molecule) roughly follows the shape of the absorption spectrum (\autoref{fig:transfer-no-cavity}a-b), as the amount of energy absorbed dictates how many carriers are excited.
However, the number of \glspl{he} injected to the molecule clearly deviates from this shape (\autoref{fig:transfer-no-cavity}c) as \gls{hc} transfer depends on generation and \emph{injection} efficiency.
The maximum injection is obtained at \qty{3.6}{\electronvolt}, which is off-resonant to the \gls{lsp}, and the number of injected \glspl{hc} is roughly constant in a range of pulse frequencies between \qty{3}{\electronvolt} and \qty{3.6}{\electronvolt}.
Thus, factoring out the efficiency of absorption (\qty{93}{\%} less energy is absorbed using the \qty{3}{\electronvolt} pulse compared to \qty{3.8}{\electronvolt}; see \autoref{sfig:cavity-energy-distances}) the injection of \glspl{he} must be much more efficient using a pulse of \qty{3}{\electronvolt}.\cite{FojRosKui22}

\textit{Level alignment --}
To understand this behavior, let us give a perturbative perspective on \gls{hc} generation.
Driving the system with an external potential $V_\text{ext}$ we perturb the density and induce a potential $\delta V$.
This is primarily the Coulomb potential of the \gls{lsp}.
\Glspl{hc} form as the \gls{lsp} starts to decay \cite{RosErhKui20}, as their coupling to $V_\text{ext}$ is much weaker than to $\delta V$.
From Fermi's golden rule we should expect a continuous drive of frequency $\omega$ to result in a rate of \gls{he} formation in state $a$ to be $1/\tau/\hbar^2\sum_{i} (2 - f_a)f_i\left|M_{ia}\right|^2\cdot\left(1/\left[(\omega - \omega_{ia})^2 + \tau^{-2}\right] + \left[(\omega + \omega_{ia})^2 + \tau^{-2}\right]\right)$ where $\hbar\omega_{ia}$ is the energy of excitation $i\rightarrow a$, $f_i$ and $f_a$ occupation numbers (including spin degeneracy of 2) and $\tau$ a characteristic lifetime of carriers \cite{ManLiuKul14}.
The denominator represents the requirement for energy-conservation during the excitation event, i.e., the longer the lifetime, the sharper the resonant condition.
\footnote{We note that in \gls{rttddft} with an adiabatic \gls{xc}-kernel, decay channels such as Auger and phonon scattering are missing and the lifetime is infinite, but because we are simulating a finite pulse with a full-width at half-maximum of \qty{0.7}{\electronvolt} in frequency space, the condition is broadened accordingly.}
The transition matrix element $M_{ia} = \int\mathrm{d}\boldsymbol{r} (V_\text{ext} + \delta V) (f_i - f_a) \psi_i^*(\boldsymbol{r})\psi_a(\boldsymbol{r})$ effectively represents the coupling strength of the \gls{lsp} to each excitation times the coupling strength of the \gls{lsp} to the driving field.

Mapping out all excitations in the system (\autoref{sfig:excitations}) we see a competition between two effects dictating the probability for an excitation to occur.
The pulse should be aligned to:
(i) the excitation energy $\omega_{ia}$ in order to satisfy the resonance condition ($\sim 1/(\omega - \omega_{ia})^2$) and
(ii) the \gls{lsp} resonance, as $\left|M_{ia}(\omega)\right|^2$ scales with the amount of energy absorbed.

The total \gls{np}+molecule system involves many energy-conserving excitations due to the dense DOS of the \gls{np} (illustrated by the accordingly colored lines in \autoref{fig:transfer-no-cavity}f), such that (i) plays a minor role and (ii) causes the \gls{he} distribution to rather uniformly decrease in amplitude when the pulse is detuned from the \gls{lsp} (\autoref{fig:transfer-no-cavity}d+b).
However, only few energy-conserving \emph{charge-transfer} excitations, i.e., from \gls{np} \emph{to} molecule, exist (\autoref{fig:transfer-no-cavity}g) due to the more discrete molecular level-structure.
For our particular \gls{np}-molecule geometry, the \gls{lumo} is hybridized with the metal, forming three distinct states at 1.73, 1.98, and \qty{2.33}{\electronvolt} above the Fermi level (\autoref{sfig:pdos}).
Across all pulse frequencies, an excitation from \gls{np} states at \qty{-0.93}{\electronvolt} to the middle orbital at \qty{1.98}{\electronvolt} ($\hbar\omega_{ia} = \qty{3.26}{\electronvolt}$) has the largest matrix element of all charge-transfer excitations (\autoref{sfig:excitations}).
The next strongest charge-transfer excitation goes from \qty{-0.93}{\electronvolt} to \qty{1.98}{\electronvolt} ($\hbar\omega_{ia} = \qty{3.02}{\electronvolt}$).
An improved alignment to these excitation energies (3.02 and \qty{3.26}{\electronvolt}) as the pulse is red-detuned causes more \glspl{he} to be injected into the middle orbital \qty{1.98}{\electronvolt} (\autoref{fig:transfer-no-cavity}f).
Competition with the alignment to the \gls{lsp} (\qty{3.8}{\electronvolt}) as well as the presence of other charge-transfer excitations with different $\omega_{ia}$ ultimately causes the plateau in \gls{he} injection between 3 and \qty{3.6}{\electronvolt}.
%
Having control over the alignment criteria (i) and (ii) presents a clear way forward to optimize \gls{hc} injection.
We begin by exploring to which extent strong coupling to electromagnetic resonator structures can be used to tune polaritonic excitations and ultimately increase injection efficiency.

\textit{Modifying the resonance with strong coupling --}
\begin{figure*}[ht]
    \includegraphics{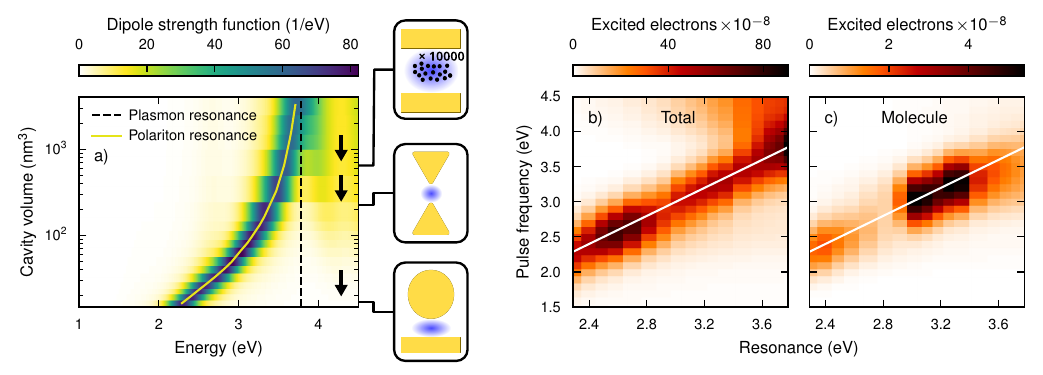}
    \caption{
    (a) Absorption spectrum of the \gls{np}-CO coupled system for different coupling strengths of the cavity.
    The spectral positions of the plasmon and the\glsfirst{lp} resonance are indicated by dashed and solid lines, respectively.
    The sketches mark effective cavity volumes that are possible to realize using (from the top) optical cavities and collective strong coupling (assuming $N=\num{10000}$), bow-tie antennas \cite{KanSchReg16}, and nanocavities \cite{BauAizMik19}.
    (b) Number of electrons excited in the system as a whole, (c) as well as in the molecule.
    The latter are plotted as a function of the spectral position of the resonance, which can be related to the cavity volume according to (a).
    }
    \label{fig:transfer-cavity}
\end{figure*}

We introduce a simplified representation of a lossy cavity mode to our \gls{np}-molecule system via the radiation-reaction potential~\cite{schafer2021shortcut}, as demonstrated in Ref.~\citenum{schaeferembedding2022}.
The frequency of the cavity mode is tuned to the \gls{lsp} resonance of \qty{3.8}{\electronvolt} and the mode features a lifetime of $\tau = \qty{17.32}{\femto\second}$, which is longer but still comparable to the lifetime of the \gls{lsp}. 
Energy is absorbed from the pulse only via the matter system to ensure comparable results for all following investigations, i.e., increasing absorption efficiency is a secondary effect of resonator structures that we do not discuss in this manuscript but that has been explored in previous experiments.
Here, we use an unspecified cavity structure to explore our hypothesis but illustrate possible realizations with typical effective cavity volumes in the insets of \autoref{fig:transfer-cavity}a.
The optical field follows Maxwell's equations which leaves the electronic Kohn-Sham orbitals, and therefore condition (i), unaffected.
Strong interaction between cavity and \gls{lsp} causes the \gls{lsp} to split in \gls{lp} and \gls{up} (\autoref{fig:transfer-cavity}a; the latter, however, is quenched at large coupling strengths due to overlapping with interband transitions).
The clue is now that the \gls{lp} can be monotonically redshifted by reducing the effective mode volume of the resonator structure or increasing the density of optical emitters inside a given volume -- we control condition (ii) by controlling the cavity.

Unsurprisingly, the total number of excited electrons in the system peaks when the system is driven in resonance with the \gls{lp} (\autoref{fig:transfer-cavity}b).
Slightly shifting the \gls{lp} approximately halves the number of electrons in the total system as the \gls{lsp} is split into two polaritonic states that carry each half of the oscillator strength.
Further increasing the cavity strength, the number of excited electrons stays roughly constant, with a notable exception between \num{2.5} and \qty{2.8}{\electronvolt}.
We attribute the increase in this region to a spectral overlap of the \gls{lp} with a feature in the spectrum at \qty{2.6}{\electronvolt} (\autoref{fig:transfer-no-cavity}a; compare spectra of larger \glspl{np} in Ref.~\citenum{RosErhKui20}).

The number of electrons \emph{injected} into the molecule does not necessarily peak at resonance, which we knew already from the no-cavity case. 
As illustrated in \autoref{fig:transfer-cavity}c, tuning the \gls{lp} into the energetic domain of optimal injection efficiency at \qty{3}{\electronvolt} results in a considerable increase of \glspl{he} injection by \qty{550}{\%} compared to the cavity-free system.
Repeating this study for other distances between the \gls{np} and molecule results in similar behavior (\autoref{sfig:summary-distances}).

\begin{figure}[ht]
    \includegraphics{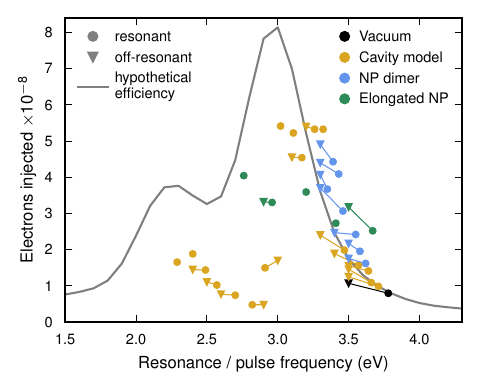}
    \caption{
    Number of electrons injected into the molecule for different setups.
    Because the principal charge-transfer excitation is red-detuned relative to the \gls{lsp} resonance, transferring the most charge requires a trade-off between being tuned to the resonance (circles) and being tuned to the charge-transfer excitation.
    Imagining that we could ``artificially'' shift the resonance yields the solid grey line.
    The slight blue-shift of the in-cavity envelope and the artificially shifted resonance can be attributed to self-polarization effects.\cite{schafer2020relevance, schafer2021shortcut}
    The data is normalized to correspond to the same absorption as the 201-atom vacuum system.
    In particular, the values for the \gls{np} dimer have been halved, and the values for the elongated \glspl{np} have been multiplied by 201 and divided by the number of atoms in each structure.
    }
    \label{fig:summary}
\end{figure}

Strong coupling allows us to tune condition (ii) on demand and reach energetic alignment between optical absorption and \gls{hc} injection.
The ideal realization of a hypothetical system, in which the \gls{lsp} can be freely tuned to match the pulse frequency, can be estimated by normalizing the injected carriers (at \qty{3.8}{\electronvolt} pulse) by the amount of energy absorbed at each pulse frequency (\autoref{fig:summary}; solid line). 
The absolute maximum in \gls{he} injection for this hypothetical system would be then obtained when the pulse frequency, the \gls{lsp} resonance, and the \qty{3}{\electronvolt} charge-transfer excitation are all resonant.
As established for our original system in vacuum (\autoref{fig:summary}; black), the real systems will not necessarily feature optimal injection for resonant (pulse vs \gls{lsp}; circles) drive but perform typically best if driven slightly off-resonantly (triangles).
Our proposed optical tuning mechanism via the \gls{lp} follows the hypothetical efficiency indeed closely down to \qty{3}{\electronvolt}.
For even larger hybridization, a significant part of the energy is trapped in the cavity subsystem (\autoref{sfig:cavity-energy-distances}, \autoref{sfig:summary-distances}) and is no longer used to generate \glspl{hc} which results in the under-performance below \qty{3}{\electronvolt}.

\textit{Atomistic cavity model --}
One possible atomistic realization of our resonator structure is to simply place a second identical \gls{np} close to the \gls{np}-molecule system.
The system is set up such that the gap of the \gls{np} dimer is between \qty{9.83}{\angstrom} (resonance \qty{3.62}{\electronvolt}; see \autoref{sfig:spectra}) and \qty{4.33}{\angstrom} (resonance \qty{3.35}{\electronvolt}) and the molecule is placed on the outer side of one of the \glspl{np}, so that charge can only be injected from that \gls{np}.
Normalizing by the increased amount of energy absorbed in this system, the number of injected electrons for this \gls{np} dimer system (\autoref{fig:summary}; blue markers) follows again the hypothetical efficiency.

\textit{Manipulation of shape --}
Popular strategies to control the optical properties of \gls{np} systems include the manipulation of shape, size, and composition \cite{KimHuyYoo23, RenTiaLou21, FojRosKum24}.
However, such approaches modify the electronic ground state which results in an overall change of optical and catalytic activity.
We construct a series of artificially elongated \glspl{np} (\autoref{fig:summary}; green markers) by inserting up to 8 atomic layers in the middle of the structure and relaxing the nuclear positions with a simple effective medium theory model (more details in SI).
The simplistic relaxation results in a shift of the \gls{lsp} and an increase in injection efficiency even for the 201 Ag \gls{np}. 
Elongating the \glspl{np}, and normalizing the injection efficiency by the increase in size, leads to a mild increase in catalytic activity.
Changing the shape modifies the ground state and with it the single-particle spectrum, moving the previously observed resonance out of reach.
Shape manipulation is clearly a valid alternative but our study illustrates that both approaches feature unique strengths and a holistic optimization strategy that accounts for shape, size, composition, and polaritonic control holds great potential to give birth to a new generation of catalytic materials.

%
%

\textit{Conclusions and outlook --}
To summarize, we have illustrated that the efficiency of the direct \gls{hc} transfer process depends on the alignment of the incoming photon energy, the \gls{lsp} resonance of the \gls{np}, and the excitation energy of a few (or even a single) charge-transfer excitations\cite{FojRosKui22}.
We have discussed a polaritonic framework for tuning the \gls{np} resonance to the charge-transfer excitations based on strong coupling to an electromagnetic resonator structure.
In this framework, we achieve a more than fivefold increase in \gls{hc} injection with the potential to fine-tune this increase non-intrusively by adjusting the configuration of the resonator.
Alternative strategies based on modifying the \gls{np} shape change \gls{lsp} and charge-transfer excitations at the same time, resulting in a more complex optimization/control problem.
Notable upsides for the use of optical environments over the adjustment of shape or composition are threefold:
(i) A chosen catalyst can be fine-tuned to a specific orbital and thus reaction, allowing for a more modular design-approach.
(ii) Some resonator structures can be used to dynamically adjust to changes of the catalyst, appearing, e.g., due to deterioration.
(iii) The established approach to boost absorption characteristics, and thus photo-catalytic activity, with optical resonators~\cite{gupta2021nanoimprint, Ren21, Jin23, HerJueKes23, NanGirSte23, Yuan2024} can be conveniently combined with our approach.

Our proposal could be validated in various mixed-plasmonic or collectively coupled systems, many of which are already in striking distance~\cite{Sarkar19, baumberg2019extreme,Hertzog21, gupta2021nanoimprint} or even exceed~\cite{mueller2020deep, herran2023plasmonic, baumberg2019extreme} the necessary coupling strength.
Minor remaining challenges are the design of resonator structures that have a sufficient quality factor and ensuring that the energy of the field is deposited into the photo-catalytic \glspl{np}.
Maxwell-TDDFT approaches, especially those following embedding concepts~\cite{schafer2021shortcut, schaeferembedding2022}, are ideal to support this task as they provide a computationally accessible framework to consistently link macroscopic optical energy distribution to microscopic \gls{hc} dynamics.
Understanding the delicate \gls{hc} injection process with the help of the adjustable non-intrusive control-knob discussed here would deepen our understanding of plasmonic catalysis and open new avenues for refined designs that will be required in the near future to elevate the hydrogen economy to the desired level.
This includes the connection between energy absorption and chemical reactivity~\cite{Corni24,gardner2023assessing} as well as the impact of prolonged non-equilibrium carrier distribution (before thermalization to \glspl{hc}) on the reactivity.

To this end, polaritonically steered plasmonic catalysis might open a path to replace precious materials, such as platinum and gold, with more abundant and optically active materials, such as aluminum~\cite{Yuan22}, thus triggering the next development step towards green chemistry.

\subsection*{Software used}
The \gpaw{} package \cite{MorHanJac05, MorLarKui24} with \gls{lcao} basis sets \cite{LarVanMor09}, \gls{lcao}-\gls{rttddft} implementation \cite{KuiSakRos15}, and radiation-reaction potential~\cite{schafer2021shortcut,schaeferembedding2022} was used for the \gls{rttddft} calculations.
The \gls{gllbsc} \cite{GriLeeLen95, KuiOjaEnk10} \gls{xc}-functional, utilizing the Libxc \cite{LehSteOli18} library, was used in \gpaw{}.
The VASP \cite{KreHaf93, KreFur96, KreFur96a, KreJou99} suite with the projector augmented wave \cite{Blo94} method and the vdW-DF-cx \cite{BerHyl14, KliBowMic09, KliBowMic11, RomSol09} \gls{xc}-functional was used for the structure relaxations.
The \textsc{ase} library \cite{LarMorBlo17} was used for constructing and manipulating atomic structures.
The NumPy \cite{HarMilvan20}, SciPy \cite{VirGomOli20}, and Matplotlib \cite{Hun07} Python packages were used for processing and plotting data.

\section*{Supporting Information}
Atomic structures, Computational details, Energetic contributions matter and cavity subsystems, (Projected) density of states, Transition contribution map, Electron injection for different \gls{np}-molecule distances, Absorption spectra.

\section*{Acknowledgments}
J.F., and P.E. acknowledge funding from the Knut and Alice Wallenberg foundation through Grant No. 2019.0140, funding from the Swedish Research Council through Grant No. 2020-04935 as well as the Swedish Foundation for Strategic Research via the SwedNESS graduate school (GSn15-0008).
C.S. acknowledges funding from the Horizon Europe research and innovation program of the European Union under the Marie Sk{\l}odowska-Curie grant agreement no.\ 101065117.
The computations were enabled by resources provided by the National Academic Infrastructure for Supercomputing in Sweden (NAISS) at NSC, PDC, and C3SE partially funded by the Swedish Research Council through grant agreement no. 2022-06725.

Partially funded by the European Union.
Views and opinions expressed are, however, those of the author(s) only and do not necessarily reflect those of the European Union or REA.
Neither the European Union nor the granting authority can be held responsible for them.

\section*{TOC Graphic}
\includegraphics{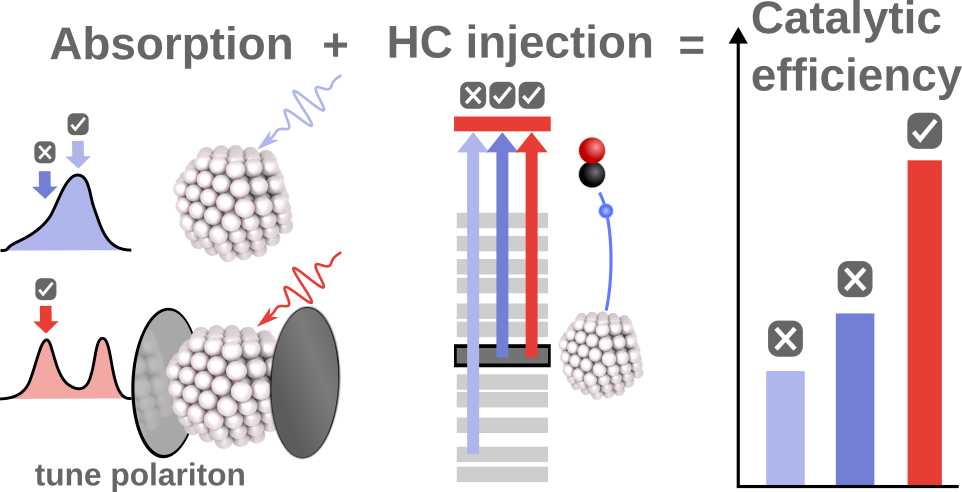}

\bibliography{references}

\end{document}


\title{Supplementary Information: Controlling Plasmonic Catalysis via Strong Coupling with Electromagnetic Resonators}

 \author{Jakub Fojt}
 \affiliation{
 Department of Physics, Chalmers University of Technology, 412 96 G\"oteborg, Sweden}

  \author{Paul Erhart}
  \affiliation{
  Department of Physics, Chalmers University of Technology, 412 96 G\"oteborg, Sweden}
  
 \author{Christian Sch\"afer}
 \email[Electronic address:\;]{christian.schaefer.physics@gmail.com}
 \affiliation{
 Department of Physics, Chalmers University of Technology, 412 96 G\"oteborg, Sweden}

\date{\today}

\maketitle

\onecolumngrid
\tableofcontents{}

\newpage

\section{Supplementary Figures}

\begin{figure}[H]
    \centering
    \includegraphics{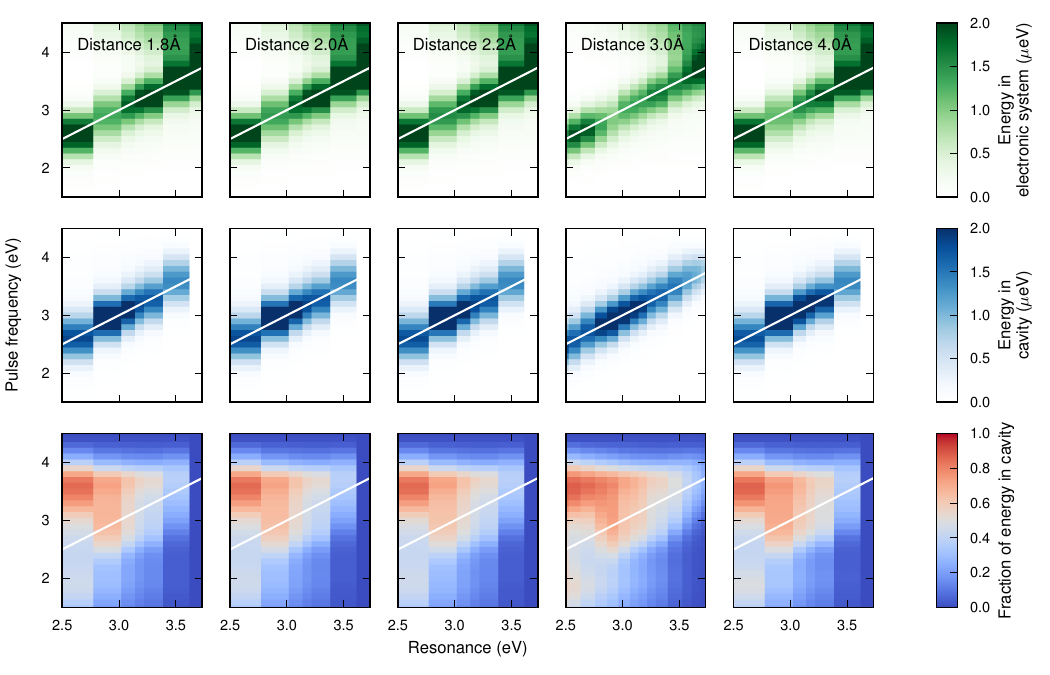}
    \titlecaptionlabel{Amount of energy in the electronic and cavity subsystems}{
    The energy supplied by the pulse is at the end of the simulation partially in the electronic subsystem and partially in the cavity subsystem.
    In the case that the resonance is redshifted away from the bare cavity resonance at \qty{3.8}{\electronvolt}, pulses resonant with the bare cavity tend to put a larger fraction into the cavity.
    Pulses above the d-band edge tend to put a larger fraction of energy into the electronic system.
    }
    {sfig:cavity-energy-distances}
\end{figure}

\begin{figure}[H]
    \centering
    \includegraphics{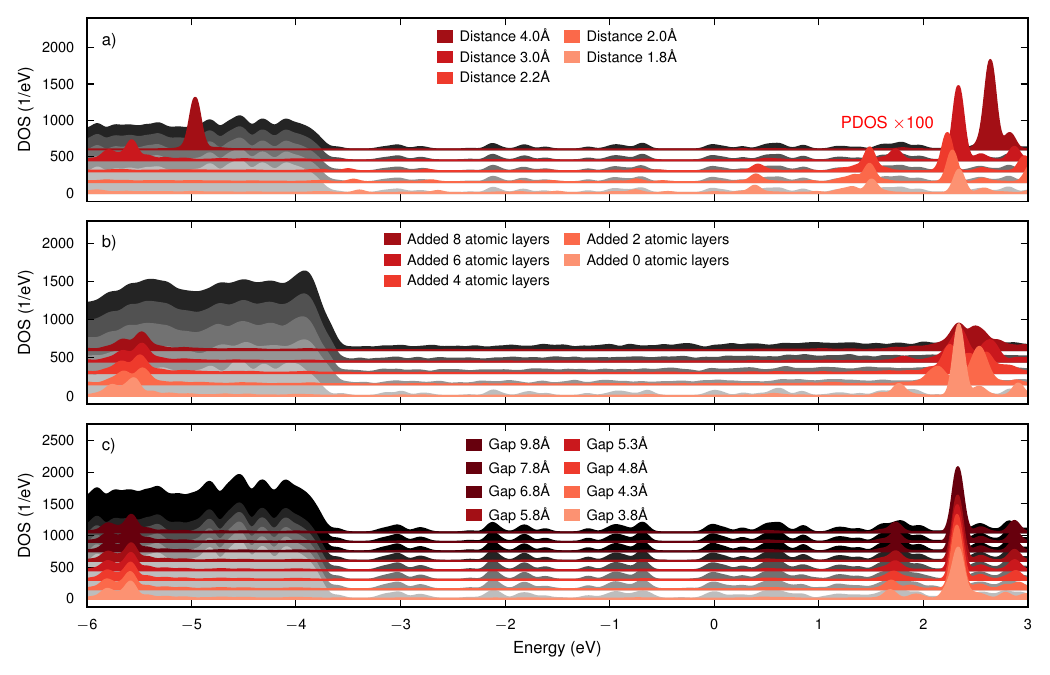}
    \titlecaptionlabel{Density of states and projected density of state of the molecule}
    {
    (a) For the NP-molecule system at various distances, (b) for the distance of \qty{3}{\angstrom} between the molecule and elongated NPs, and (c) for the two NP, one molecule systems.
    For distances closer than \qty{3}{\angstrom} the molecule is strongly hybridized with the NP, splitting the original LUMO state into many.
    Elongating the NP by adding several atomic layers to it also affects the hybridization, despite the molecule being kept at the same distance from the surface.
    Varying the gap in the NP dimer keeps the ground state unchanged, except for at the very smallest gaps, where tiny shifts in the DOS and PDOS are visible.} 
    {sfig:pdos}
\end{figure}

\begin{figure}[H]
    \centering
    \includegraphics{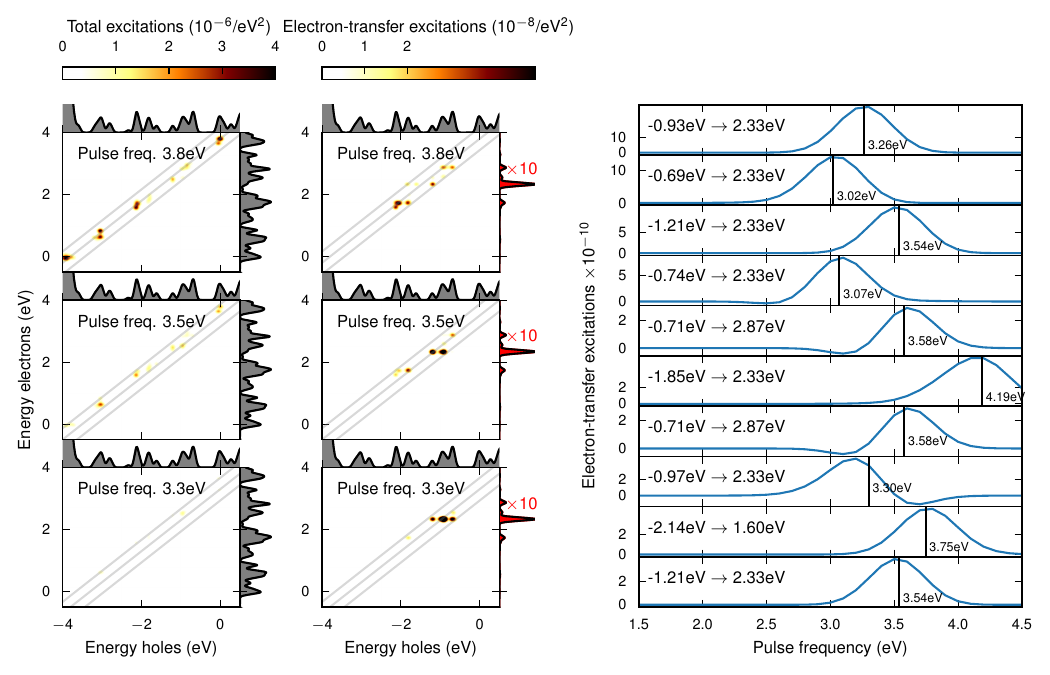}
    \titlecaptionlabel{Transition contribution map for the \qty{3}{\angstrom} distance, no cavity case.}{The contribution map to total excitations (NP and molecule), as well as projected on the molecule are shown after excitation with three different pulse frequencies.
    In the rightmost panels, the 10 strongest electron-transfer excitations (that is contribution to the projected transition contribution map) for all pulses between \qty{1.5}{\electronvolt} and \qty{4.5}{\electronvolt} are shown.
    Transitions between degenerate states have been summed and plotted together.
    The strongest transition is one from the \qty{-0.93}{\electronvolt} NP state to the \qty{2.33}{\electronvolt} molecular state, which contributes the most to the total charge transfered when the pulse frequency is resonant with the transition energy of \qty{3.26}{\electronvolt}.
    }
    {sfig:excitations}
\end{figure}

\begin{figure}[H]
    \centering
    \includegraphics{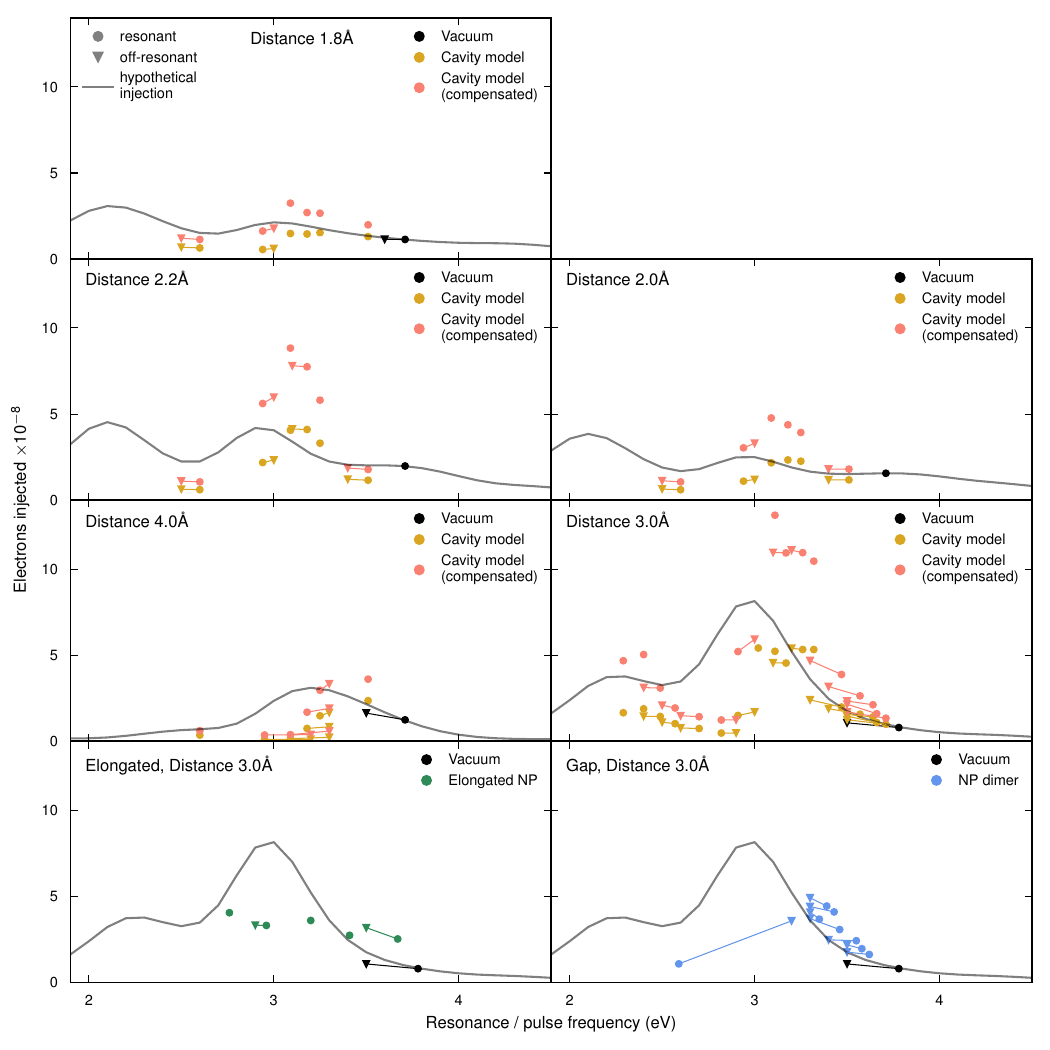}
    \titlecaptionlabel{Number of electrons injected for all considered systems}{
    Collection of results for different NP-molecule distances, elongated NPs, and NP dimers presented consistent with Fig.~3 of the main manuscript. Elongated and NP dimer efficiencies are normalized such that the same amount of energy is absorbed (see main manuscript) to ensure a fair comparison. As detailed in Fig.~\ref{sfig:pdos} and mentioned in the main manuscript, the cavity acts as as energy deposit such that only a fraction of the absorbed energy can be used to generate HCs. Compensating for this 'lost' energy by scaling the efficiency by the energy missing in the matter system ("Cavity model (compensated)"; orange) pushes the efficiency even beyond the hypothetical curve. This suggests that the LSP changes its character when strongly coupled to the cavity, emphasizing the need for reliable theoretical investigations that go beyond optimization based on the simplistic hypothetical curve. Very small distances between the NP dimers results in strong hybridization which in turn strongly modifies the LSP (see Fig.~\ref{sfig:spectra}) and leads to the outlier at 2.6~eV resonance energy.
    }
    {sfig:summary-distances}
\end{figure}

\begin{figure}[H]
    \centering
    \includegraphics{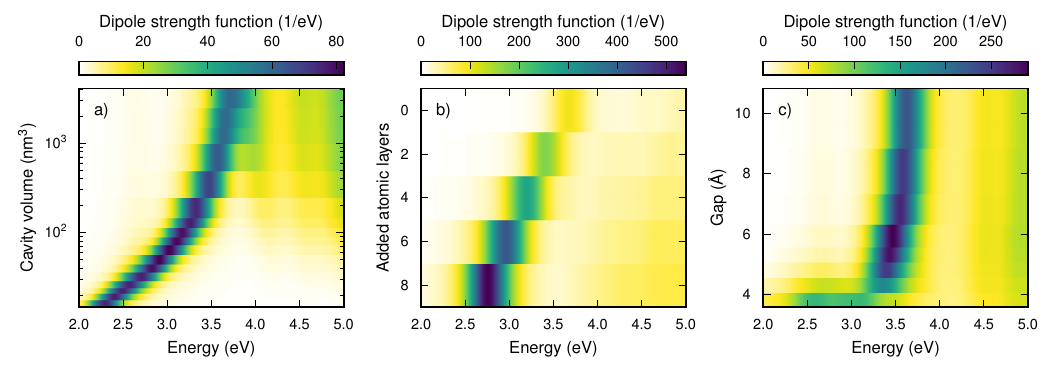}
    \titlecaptionlabel{Absorption spectra}{(a) For the \qty{3}{\angstrom} distance system in an optical cavity. The coupling strength to the cavity is inversely proportional to the cavity volume.
    (b) For the NPs which have been elongated by inserting atomic layers in the middle.
    As the number of atomic layers is added, the number of electron in the system, and thus total  absorption increases.
    (c) For the NP dimer. For the smallest gap, the spectrum is drastically broadened.}
    {sfig:spectra}
\end{figure}

\newpage
\twocolumngrid
\section{Supplementary Methods}

\subsection{Atomic structures}

The atomic structure for the 201-atom Ag NP was taken from from Ref.~\citenum{FojRosKui22}.
This NP has the shape of a regular truncated octahedron.
In Ref.~\citenum{FojRosKui22}, the NP was relaxed with VASP \cite{KreHaf93, KreFur96, KreFur96a} using a plane-wave basis set, the PAW \cite{Blo94, KreJou99} method, the vdW-df-cx \cite{DioRydSch04, BerHyl14, KliBowMic09, RomSol09} XC-functional, a plane wave cutoff \qty{500}{\electronvolt}, and Gaussian occupation number smearing scheme with the parameter \qty{0.1}{\electronvolt}.
Structure relaxations were performed using the conjugate gradient relaxation method implemented in VASP.
Relaxation was stopped when the maximal force on any atom fell below \SI{0.015}{\electronvolt\per\angstrom}.

For the analysis in the main text, we place a CO molecule with the bond length \qty{1.144}{\angstrom} at a distance of \qty{3}{\angstrom} from the center atom of the (111) face of the NP.
The molecule is oriented with the C atom towards the surface, and the bond axis perpendicular to the surface.
The polarization of the applied field in the TDDFT simulation is chosen along the bond axis.
After constructing the NP+molecule structure, it \textit{is not} relaxed further, in order to study the charge transfer at a fixed distance.
In the SI, we also present results using structures where the distances are 1.8, 2.0, 2.4 and \qty{4}{\angstrom}.

For the analysis in the main text, we construct a NP dimer by duplicating the NP and shifting it a distance along the bond axis of the molecule.
The CO molecule is added at a distance of \qty{3}{\angstrom} in the same configuration as before, in such a way that it is on the outer side of the NP dimer.
We shift the NP so the center-center distance is between 24 and \qty{18.5}{\angstrom}, corresponding to an edge-to-edge distance of between 
9.83 and \qty{4.33}{\angstrom}.
In the SI, we also present an additional data point where the center-center distance is \qty{18}{\angstrom} and the edge-to-edge distance \qty{3.83}{\angstrom}.

For the analysis in the main text, we construct a series of artificially elongated NPs, by splitting the NP in the middle, shifting the structures apart to make room for two new layers, and duplicating the two atomic layers in the middle of the NP and placing them in the gap. To preserve the surfaces of the NP, the shifted half also has to be shifted laterally, in order to create a tilted structure.
This is repeated between 0 and 4 times to create structures with between 0 and 8 added layers.
These structures are relaxed with the EMT method and BFGS optimizer implemented in ASE \cite{LarMorBlo17} until the maximal force is \SI{0.001}{\electronvolt\per\angstrom}.

\subsection{Computational details}

We carried out DFT and real-time TDDFT calculations using the \gpaw{} package \cite{MorHanJac05, MorLarKui24} with LCAO basis sets \cite{LarVanMor09}, LCAO-RTTDDFT implementation \cite{KuiSakRos15}, and the \textsc{ase} library \cite{LarMorBlo17}.

For the computation of the ground state we used the GLLB-SC \cite{GriLeeLen95, KuiOjaEnk10} XC-functional, utilizing the Libxc \cite{LehSteOli18} library.
We used the \emph{pvalence} \cite{KuiSakRos15} basis set for Ag, which is optimized to represent bound unoccupied states, and the dzp basis set for CO.
The structures were padded with at least \qty{6}{\angstrom} of vacuum, and we used a grid spacing of \qty{0.2}{\angstrom} for wave functions and \qty{0.1}{\angstrom} for potentials.
The Coulomb potential was represented in numerical form on the grid, with an additional analytic moment correction \cite{CasRubSto03} centered at the NP.
Fermi-Dirac occupation number smearing with width \qty{0.05}{\electronvolt} was used.
The self-consistent loop was stopped when the integral of the difference between two subsequent densities was less than \qty{1e-12}{}.
Pulay \cite{Pul80}-mixing was used to accelerate the ground state convergence.

We applied a \deltakick{} \cite{YabBer96} of strength $10^{-5}$ in atomic units, and performed time propagation in steps of \qty{10}{\atto\second} for a total length of \qty{30}{\femto\second} using the adiabatic GLLB-SC kernel.
The Fourier transform of the Kohn-Sham density matrix was built up on the fly during propagation and saved on a predefined frequency grid for later analysis.
The optical cavity was included in the simulations using the radiation-reaction potential~\cite{schafer2021shortcut,schaeferembedding2022}, implemented in a publicly available branch of GPAW \footnote{Public GPAW fork \url{https://gitlab.com/christian.schaefer.physics/gpaw/-/tree/qed/}. Please note that the implementation is experimental and the API is subject to change}.

For the computation of HCs, including projections on the molecule, we used the method of Refs.~\citenum{FojRosKui22, RosErhKui20}. Details are found in the reference.
In short, we post process the saved Fourier transform the the KS density matrices to calculate the response to a Gaussian pulse
\begin{align}
    \label{eq:laser}
    \mathcal{E}_z(t) = \mathcal{E}_0 \cos(\omega_0 t) \exp(- (t-t_0)^2/\tau_0^2)
\end{align}
of frequency $\omega$, strength $\mathcal{E}_0 =  \qty{51}{\micro\volt\per\angstrom}$, peak time $t_0 = \qty{10}{\femto\second}$, and duration $\tau_0 = \qty{2.1}{\femto\second}$ (corresponding to a FWHM in frequency space of \qty{0.7}{\electronvolt}).

We computed the total density of states as
\begin{align}
    \sum_k \delta(\varepsilon - \varepsilon_k)
\end{align}
and the PDOS for the molecule as
\begin{align}
    \sum_k \delta(\varepsilon - \varepsilon_k) \int_\text{mol} \left|\phi^{(0)}_k(\vec{r})\right|^2 \dd{\vec{r}},
\end{align}
where $\varepsilon_k$ and $\phi^{(0)}_k(\vec{r})$ are the KS eigenvalues and wave functions, and the integral goes over the Voronoi region of the molecule (that is each point in space which is closer to the molecule than any other atom).
For visualization, the $\delta$-functions in energy were replaced by a Gaussian $(2\pi\sigma^2)^{-1/2} \exp(-\varepsilon^2/2\sigma^2)$ with width $\sigma=\qty{0.07}{\electronvolt}$.

\bibliography{references}